\documentclass[twocolumn,showkeys,showpacs,preprintnumbers,prd,superscriptaddress,nofootinbib]{revtex4-1}
\usepackage{graphicx}
\usepackage{epsf}
\usepackage{bm}
\usepackage{amsmath}
\usepackage{amsfonts}
\usepackage{amssymb}
\usepackage{epstopdf}
\usepackage{natbib}
\usepackage{hyperref}
\usepackage{color}
\usepackage{verbatim}
\usepackage{multirow}
\usepackage{bm}
\usepackage{hyperref}

\makeatletter\let\expandableinput\@@input\makeatother

\newcommand{\nq}[1]{%
	\begin{tabular}{@{}c@{}}\strut#1\strut\end{tabular}%
}

\begin{document}

\title{Exploring the $H_0$ tension and the evidence of dark sector interaction\\ from 2D BAO measurements}

\author{Armando Bernui}
\email{bernui@on.br}
\affiliation{Observat\'orio Nacional, 
Rua General Jos\'e Cristino 77, S\~ao Crist\'ov\~ao, 20921-400, Rio de Janeiro, RJ, Brazil}

\author{Eleonora Di Valentino}
\email{e.divalentino@sheffield.ac.uk}
\affiliation{School of Mathematics and Statistics, University of Sheffield, Hounsfield Road, Sheffield S3 7RH, United Kingdom}

\author{William Giar\`e}
\email{w.giare@sheffield.ac.uk}
\affiliation{School of Mathematics and Statistics, University of Sheffield, Hounsfield Road, Sheffield S3 7RH, United Kingdom}

\author{Suresh Kumar}
\email{suresh.math@igu.ac.in}
\affiliation{Department of Mathematics, Indira Gandhi University, Meerpur, Haryana 122502, India}

\author{Rafael C. Nunes}
\email{rafadcnunes@gmail.com}
\affiliation{Instituto de F\'{i}sica, Universidade Federal do Rio Grande do Sul, 91501-970 Porto Alegre RS, Brazil}
\affiliation{Divis\~{a}o de Astrof\'{i}sica, Instituto Nacional de Pesquisas Espaciais, Avenida dos Astronautas 1758, S\~{a}o Jos\'{e} dos Campos, 12227-010, S\~{a}o Paulo, Brazil}

\begin{abstract}
We explore observational constraints on a cosmological model with an interaction between dark energy (DE) and dark matter (DM), using a compilation of 15 measurements of the 2D BAO (i.e., transversal) scale in combination with Planck-CMB data, to explore the parametric space of a class of interacting DE models.  We find that 2D BAO measurements can generate different observational constraints compared to the traditional approach of studying the matter clustering in the 3D BAO measurements.  Contrary to the observations for the $\Lambda$CDM and IDE models when analyzed with Planck-CMB + 3D BAO data, we note that Planck-CMB + 2D BAO data favor high values of the Hubble constant $H_0$.  From the joint analysis with Planck-CMB + 2D BAO + Gaussian prior on $H_0$, we find $H_0 = 73.4 \pm 0.88$ km/s/Mpc. We conclude that the $H_0$ tension is solved in the IDE model with strong statistical evidence (more than 3$\sigma$) for the IDE cosmologies.
\end{abstract}

\keywords{}

\pacs{}

\maketitle

\section{Introduction}
\label{sec:introduction}

The standard cosmological model (the $\Lambda$CDM-cosmology framework) has consolidated as the best scenario able to fit the astronomical observations carried out over the past two decades, but recently some tensions and anomalies have turned out to be statistically significant while analyzing different data sets, placing the $\Lambda$CDM cosmology in a crossroad (see~\cite{Abdalla:2022yfr} for a review). The most statistically significant tension in the literature is in the determination of the Hubble constant, $H_0$, between the Cosmic Microwave Background (CMB) and the direct local distance ladder measurements. Assuming the $\Lambda$CDM scenario and the Planck-CMB data~\cite{Planck:2018vyg}, these observations are at more than 5$\sigma$ tension with the SH0ES measurement~\cite{Riess:2021jrx,Riess:2022mme}. And, in general, while the early time estimates prefer a lower value for $H_0$, the late time measurements are in agreement with a higher value (see~\cite{Abdalla:2022yfr} for a review). The lower value of $H_0$ inferred from the Planck-CMB data is actually in very good agreement with Baryon Acoustic Oscillations (BAO) + Big Bang Nucleosynthesis (BBN) constraints~\cite{Cuceu:2019for,Schoneberg:2019wmt,Schoneberg:2022ggi}, and the other CMB experiments like ACTPolDR4 and SPT-3G~\cite{Aiola_2020,Dutcher_2021}. Motivated by several current observational discrepancies, unlikely to disappear completely by introducing multiple and unrelated systematic errors, it has been widely discussed in the literature whether new physics beyond the $\Lambda$CDM can solve these tensions, in particular the $H_0$ tension~\cite{DiValentino:2021izs,Perivolaropoulos:2021jda,Schoneberg:2021qvd}. One of the most popular dark energy (DE) models in literature, designated by interacting DE (IDE) models~\cite{Wang_2016}, where a non-gravitational interaction between the dark matter (DM) and DE is postulated to exist, have been intensively investigated as a possibility to solve the current cosmological $H_0$ tension~\cite{Yang:2021hxg,Gao:2021xnk,Yao:2020pji,Martinelli:2019dau,Lucca:2020zjb,Yang:2019uog,DiValentino:2019ffd,Pan:2019jqh,Kumar:2019wfs,Yang:2018euj, Kumar:2016zpg, Murgia:2016ccp, Kumar:2017dnp, DiValentino:2017iww, Yang:2018ubt, Yang:2019uzo, Pan:2019gop, DiValentino:2019jae, DiValentino:2020leo, DiValentino:2020kpf, Amjad_h0, Gomez-Valent:2020mqn, Yang:2020uga, Yao:2020hkw, Pan:2020bur, DiValentino:2020vnx, Kumar:2021eev,Yang:2021oxc,Lucca:2021eqy,Halder:2021jiv,Gariazzo:2021qtg,Nunes:2021zzi,Akarsu:2021fol,Akarsu:2022typ,Yang:2022csz}.\footnote{It has also been investigated where the IDE can solve/alleviate the $S_8$ tension~\cite{Pourtsidou:2016ico,Lucca:2021dxo,Poulin_S8,Naidoo_qw}.} On the other hand, it is also discussed where in fact is the ability of this class of models in actually solving the $H_0$ tension using robust external probes other than Type Ia supernovae (SN) and BAO sample~\cite{Nunes:2022bhn}, or even the non-ability for late times modification in solving the $H_0$ problem~\cite{Camarena:2021jlr,Efstathiou_2021,Lemos_2018,Cai_2022}.

The state-of-the-art constraints on IDE cosmologies arise
primarily from CMB data in combination with 
measurements of the late-time background expansion history from BAO and SN. On the other hand, the traditional approach to study the matter clustering in the three-dimensional (3D) BAO phenomena is to assume a fiducial cosmology 
to calculate the 3D comoving distances, then perform the 2-point correlation function from which one estimates the sound horizon scale at the end of the baryon drag epoch, $r_{s}$, the spherically averaged distance $D_V$, and as sub-products the Hubble parameter $H$, and the angular diameter distance $D_A$. 
This means that 3D BAO analyses are model dependent and their results --to some extent-- too; not to mention that usually, 
the assumed fiducial cosmology is flat-$\Lambda$CDM or type-$\Lambda$CDM. The way to investigate the impact of this assumption is to consider other fiducial cosmologies, but these consistency analyses do not include, e.g., non-flat geometries or non-$\Lambda$CDM models (e.g., extended DE cosmologies like IDE, which may differ significantly from the $\Lambda$CDM model). 

One can study the BAO phenomena without assuming a fiducial cosmology, this can be done with the transversal BAO, termed 2D BAO. 
An additional advantage is that the examination of 2D BAO phenomena does not assume a 3D geometry because it works with data 
on spherical shells, with redshift thickness $\Delta z$, and consider only their angular distribution. 
The redshift shell, where data analysis is displayed, cannot be too thick (because the projection effect smooths and shifts the 
BAO signal\footnote{The projection effect, present in any shell with $\Delta z \ne 0$, introduces a shift in the 2D BAO 
signal that is corrected using cosmological model; at the end one verifies that the shift correction is a small fraction 
of the error measurement --for a large set of models/parameters~\citep{Sanchez11}-- and for this the measurement is weakly dependent on a fiducial cosmological model. Also, smaller is $\Delta z$, smaller is the shift.}) 
neither too thin (because the number density of data should be enough to obtain a good BAO signal to noise). 
Given a deep astronomical survey with large surveyed area, one divides the volume data in several disjoint (to avoid correlation 
between contiguous shells) redshift shells. 
The analyses of these shells provide the angular scale of the 2D BAO signal  $\theta_{\text{BAO}}(z)$ at redshift $z$, 
or equivalently give measurements of $D_A(z)$ with $r_{s}$ as a parameter. Given the weakly model-dependent features of 2D BAO measurements, it seems natural to use $\{\theta_{\text{BAO}}(z)\}$ data to test non-$\Lambda$CDM models, and to perform comparisons between analyses with diverse data sets.

Distinctive features of 2D BAO measurements that make them potentially useful for studying non-$\Lambda$CDM models include: 
\begin{enumerate}
\item Important systematic effects in 3D BAO, like Redshift Space Distortions, are absent in thin redshift bins 2D BAO measurements; 
\item While 3D BAO measures $D_V(z)$, which is a combination of $H$ and $D_{\!A}$ distances, additional datasets are required to finally obtain $H(z)$ or $D_{\!A}(z)$; 
\item 3D BAO requires a passive cosmic tracer (i.e., cosmic objects that do not evolve in the large 3D volume in analysis), and in practice, a few measurements of  $\{D_V(z_i)\}$ are expected along the universe history. 
On the other hand, 2D BAO can be done with diverse cosmic tracers, in many thin redshift bins. 
\item Although 3D BAO have smaller error bars than the 2D case, this is a consequence of the use of a fiducial cosmology, and ultimately, one has to decide whether to use data with model-dependent $\sim 1\%$ 
errors, or measurements with weakly model-dependent $\sim 10\%$ errors. 
In this scenario, it is advisable to use caution and prefer the use of 2D BAO data to examine non-$\Lambda$CDM models such as IDE.
\end{enumerate}
Overall, these factors lead us to argue that 2D BAO data are suitable to provide 
remarkable observational constraints in non-$\Lambda$CDM models. 

The aim of this \textit{Letter} is to investigate the impact of the $\theta_{\text{BAO}}(z)$ measurements on a class of IDE models well studied in the literature, where in a homogeneous and isotropic universe, the dark interaction is quantified by

\begin{eqnarray}
\label{DE_DM_1}
\nabla_{\mu}T_{i}^{\mu \nu }=Q_{i}^{\nu}\,, \quad \sum\limits_{i}{%
Q_{i}^{\mu }}=0~.
\end{eqnarray}
Here the index $i$ runs over DM and DE, and the four-vector $Q_{i}^{\mu}$ governs the interaction. 

In the present work, we consider a very well known parametric form of the interaction function $Q$, namely, $Q = \mathcal{H} \,\xi \,\rho_x$, where $\xi$ is the coupling parameter between the dark components. All model formalism, the background cosmic evolution, and the DM and DE density perturbation mode evolution are well described in~\cite{Gavela_2010}. Also, as well described in previous works, in order to avoid early times instabilities, in developing the results of this work, we fix the equation of state of DE to $w = -0.9999$, and impose the coupling parameter to be negative, that is, $\xi < 0$. This condition corresponds to the energy flow from DM to DE. 

This work is structured as follows. In Section~\ref{sec:data} we present the data sets and methodology used in this work. In Section~\ref{sec:res} we discuss the main results of our analysis. In Section~\ref{sec:conclusions} we outline our final considerations and perspectives.

\section{Data sets and Methodology}
\label{sec:data}

We describe below the observational data sets and the
statistical methods we use to explore our parameter space.

\begin{itemize}

\item CMB: Measurements of CMB temperature and polarization power spectra, as well as their cross-spectrum, from the \textit{Planck} 2018 legacy data release. We consider the high-$\ell$ \texttt{Plik} likelihood for TT (in the multipole range $30 \leq \ell \leq 2508$) as well as TE and EE (in the multipole range $30 \leq \ell \leq 1996$), in combination with the low-$\ell$ TT-only ($2 \leq \ell \leq 29$) likelihood based on the \texttt{Commander} component-separation algorithm in pixel space, as well as the low-$\ell$ EE-only ($2 \leq \ell \leq 29$) \texttt{SimAll} likelihood~\cite{Planck:2019nip}. We refer to this data set as \textit{Planck}.

\item CMB lensing: Measurements of the CMB lensing power spectrum as reconstructed from the temperature 4-point correlation function~\cite{CMB_lens_2020}. We refer to this data set as \textit{lensing}.

\item BAO: Baryon Acoustic Oscillation (BAO) distance and expansion rate measurements from the final measurements of the SDSS collaboration covering eight distinct redshift intervals, obtained
and improved over the past 20 years~\cite{Alam_2021}. These consists of isotropic BAO measurements of $D_V(z)/r_s$ (with $D_V(z)$ and $r_s$ the spherically averaged volume distance, and sound horizon at baryon drag respectively) and anisotropic BAO measurements of $D_M(z)/r_s$ and $D_H(z)/r_s$ (with $D_M(z)$ the comoving angular diameter distance and $D_H(z)=c/H(z)$ the Hubble distance). All these measurements are compiled in Table 3 in~\cite{Alam_2021}, regarding BAO-only data. We refer to this data set as \textit{BAO}.

\item Transversal BAO: 
Measurements of BAO 2D, $\theta_{\text{BAO}}(z)$, obtained in a weakly model-dependent approach, compiled in table I in~\cite{Nunes_2020_theta_bao,decarvalho}. 
These measurements were obtained using public data releases (DR) of the Sloan Digital Sky Survey (SDSS), namely: DR7, DR10, DR11, DR12, DR12Q (quasars), 
and following the same methodology. 
It is important to notice that due to the cosmological-independent methodology, these transversal BAO measurements have their errors larger than the errors obtained using a fiducial cosmology. 
The reason for this fact is that, while in the former methodology the error is given by the measure of how large is the BAO bump, in the latter approach the model-dependent best-fit of the BAO signal quantifies a smaller error. 
Typically, in the former methodology the error can be of the order of $\sim 10\%$, but in some cases it can arrive to $\sim 18\%$, and in the later approach it is of the order of few percent~\cite{Sanchez11}.
Another important feature of these 2D BAO data is that the measurements at different $z$ are not correlated. 
In fact, the methodology adopted to perform these measurements excluded the possibility for covariance between measurements because the analyses of the 2-point angular correlation function were done with cosmic objects belonging to disjoint redshift shells to avoid correlation between contiguous data bins. 
Notice also that the production of the random sets, necessary to the 2PACF analyses, is also done in a model-independent way: 
consider a data sample of cosmic objects in a thin redshift bin, then one shifts randomly the angular coordinates of each object obtaining a set with the same number of cosmic objects as the data sample with randomized positions: {\it the random set} 
(see the Appendix B in~\cite{deCarvalho18}); 
clearly these random sets are subjected to the null-test to see if they are really random and no clustering signal is inside. We refer this data set to as \textit{BAOtr}.

\item SH0ES. A gaussian prior on the Hubble constant as measured by the SH0ES collaboration~\cite{Riess:2021jrx}, \emph{i.e.}, $H_0 = (73.04 \pm 1.04)$ km/s/Mpc.  We refer to this data set as \textit{$H_0$}.

\end{itemize}

We use \texttt{CLASS+MontePython} code~\cite{Lesgourgues:2011re,Audren:2012wb,Brinckmann:2018cvx} with Metropolis-Hastings mode to derive constraints on cosmological parameters for the IDE model baseline from several combinations of the data sets defined above, ensuring a Gelman-Rubin convergence criterion of $R - 1 < 10^{-2}$ in all the runs.  In what follows, we discuss the main results of our analyses. 

For the purpose of model comparison, we compute the Bayes factor $\ln B_{ij}$ to estimate the Bayesian Evidence of the IDE model with respect to the $\Lambda$CDM scenario, through the publicly available package \texttt{MCEvidence},\footnote{\href{https://github.com/yabebalFantaye/MCEvidence}{github.com/yabebalFantaye/MCEvidence}~\cite{Heavens:2017hkr,Heavens:2017afc}.}, and report the results in the last row of Table~\ref{tab.IDE}. We use the convention of a negative value if the IDE model is preferred against the $\Lambda$CDM scenario, or vice versa, and we refer to the revised Jeffrey's scale by Trotta~\cite{Kass:1995loi,Trotta:2008qt}, to interpret the results. We will say that the evidence is inconclusive if $0 \leq | \ln B_{ij}|  < 1$, weak if $1 \leq | \ln B_{ij}|  < 2.5$, moderate if $2.5 \leq | \ln B_{ij}|  < 5$, strong if $5 \leq | \ln B_{ij}|  < 10$, and very strong if $| \ln B_{ij} | \geq 10$.

\section{Results}
\label{sec:res}

\begin{table*}
\caption{Constraints at 68\% (95\%) [99\%] CL on selected parameters of the IDE model obtained from various data set combinations as indicated in the upper section of the Table. Bayes factors $\mathcal{B}_{ij}$ given by  $\ln \mathcal{B}_{ij} = \ln \mathcal{Z}_{\rm LCDM} - \ln \mathcal{Z}_{\rm IDE}$, are also displayed for the different analyses, so that a negative value indicates a preference for the IDE model against the $\Lambda$CDM scenario. }
\begin{center}
\renewcommand{\arraystretch}{2}
\resizebox{\textwidth}{!}{
\begin{tabular}{l | c c c c c c c c c c c c c c c }
\hline
\textbf{Parameter} &  \textbf{Planck} & \textbf{\nq{\\ + lensing}} & \textbf{Planck + BAO} &
\textbf{\nq{\\ + lensing}} & \textbf{Planck + BAOtr} &
\textbf{\nq{\\ + lensing}} & \textbf{Planck + BAOtr + $H_0$} &
\textbf{\nq{\\ + lensing}} \\ 
\hline\hline

{\boldmath$\xi $} 
&$ -0.43^{+0.28}_{-0.21} $  & $-0.40^{+0.23}_{-0.20}$ & $> -0.207$ & $> -0.210$ & $-0.683^{+0.088}_{-0.11}$ & $-0.683^{+0.087}_{-0.12}$ & $-0.58\pm 0.11$ & $-0.53\pm 0.11$\\
{} &$(> -0.775)$ & $(-0.40^{+0.40}_{-0.32})$ & $(> -0.389)$ & $(> -0.411)$ & $(-0.68^{+0.21}_{-0.19})$ & $(-0.68^{+0.23}_{-0.20})$ & $(-0.58^{+0.22}_{-0.21})$ & $(-0.53^{+0.19}_{-0.20})$ \\
{} &$[> -0.819]$ & $[> -0.743]$ & $[> -0.486]$ & $[> -0.527]$ & $[-0.68^{+0.29}_{-0.23}]$ & $[-0.68^{+0.37}_{-0.27}]$ & $[-0.58^{+0.31}_{-0.29}]$ & $[-0.53^{+0.39}_{-0.25}]$ \\

{\boldmath$H_0 $ \bf [Km/s/Mpc]} & $71.7^{+2.3}_{-2.7}$ & $71.6\pm 2.1$ & $68.93^{+0.79}_{-1.2}$ & $69.08^{+0.74}_{-1.3}$ & $75.2^{+1.2}_{-0.75}$ & $75.3^{+1.3}_{-0.75}$ & $73.99 \pm 0.88$ & $73.45^{+0.71}_{-0.59}$ \\

{\boldmath$S_{8 }        $} & $1.109^{+0.063}_{-0.28}$ & $1.053^{+0.079}_{-0.21}$ & $0.891^{+0.025}_{-0.062}$ & $0.893^{+0.021}_{-0.065}$ & $1.49^{+0.24}_{-0.29}$ & $1.49\pm 0.26$ & $1.23^{+0.11}_{-0.22}$ & $1.15^{+0.10}_{-0.14}$\\

{\boldmath$r_{s }$ \bf [Mpc]} & $147.08\pm 0.30$ & $147.12\pm 0.27$ & $147.03\pm 0.25$ & $147.05\pm 0.25$ & $147.32\pm 0.27$ & $147.35\pm 0.29$ & $147.31^{+0.25}_{-0.29}$ & $147.32^{+0.26}_{-0.29}$ \\

\hline

{\boldmath$\ln B_{ij}$ } &  $0.85$ & $-0.17$ & $1.60$ & $0.60$ & $-9.22$ & $-11.68$ & $-14.04$ & $-15.21$\\

\hline \hline
\end{tabular} }
\end{center}

\label{tab.IDE}
\end{table*}

\begin{table*}
\caption{Constraints at 68\% CL on selected parameters of the $\Lambda$CDM model obtained from various data set combinations.}
\begin{center}
\renewcommand{\arraystretch}{2}
\resizebox{\textwidth}{!}{
\begin{tabular}{l | c c c c c c c c c c c c c c c }
\hline
\textbf{Parameter} &  \textbf{Planck} & \textbf{\nq{\\ + lensing}} & \textbf{Planck + BAO} &
\textbf{\nq{\\ + lensing}} & \textbf{Planck + BAOtr} &
\textbf{\nq{\\ + lensing}} & \textbf{Planck + BAOtr + $H_0$} &
\textbf{\nq{\\ + lensing}} \\ 
\hline\hline

{\boldmath$H_0$ \bf [Km/s/Mpc]} & $67.32\pm 0.62$ & $67.32\pm 0.53$ & $67.65\pm 0.44$ & $67.60\pm 0.43$ & $69.01\pm 0.51 $ & $68.85\pm 0.55$ & $69.88\pm 0.48$ & $69.65\pm 0.44$ \\

{\boldmath$S_{8 }        $} & $0.832\pm 0.016$ & $0.834\pm 0.013$ & $0.825\pm 0.012$ & $0.827\pm 0.011$ & $0.794\pm 0.013$ & $0.802\pm 0.012$ & $0.774\pm 0.013$ & $0.7871^{+0.0095}_{-0.011}$\\

{\boldmath$r_{s }   $ \bf [Mpc]} & $147.06\pm 0.30$ & $147.04\pm 0.27$ & $147.21^{+0.23}_{-0.26} $ & $147.13\pm 0.23$ & $147.75\pm 0.26$ & $147.64\pm 0.26$ & $148.06\pm 0.25$ & $147.91\pm 0.24$ \\

\hline

\hline \hline
\end{tabular} }
\end{center}
\label{tab.LCDM}
\end{table*}

\begin{figure}[hbt!]
      \includegraphics[width =0.5 \textwidth]{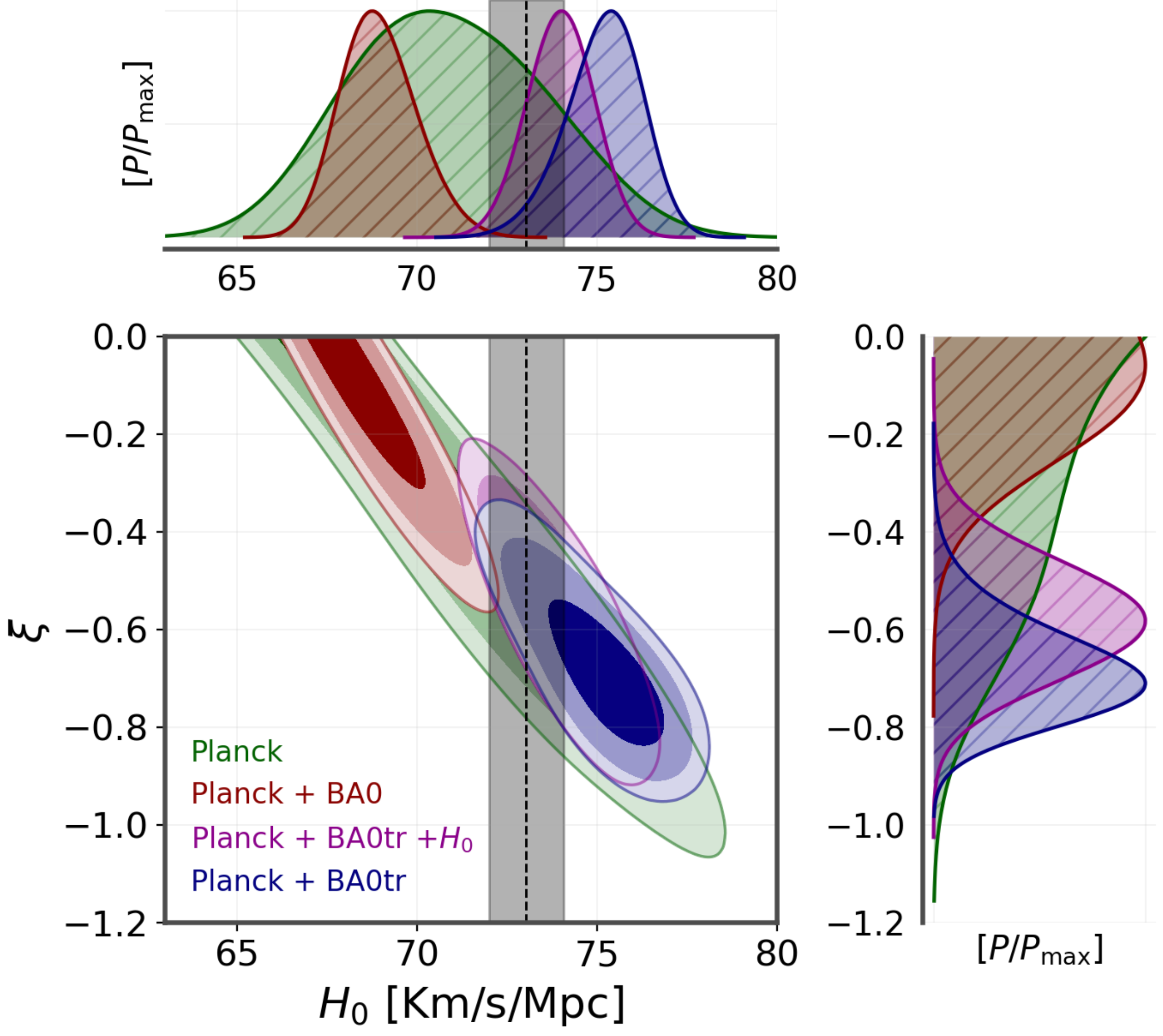} 
   \caption{2D contours at 68\%, 95\%, and 99\% CL and 1D posteriors for the cases without lensing only. The grey vertical region refers to the value of $H_0$ measured by the SH0ES collaboration ($H_0=73.04 \pm 1.04$ km/s/Mpc at 68\% CL).} 
    \label{fig:PS_IDE}
\end{figure}

\begin{figure*}
\includegraphics[width=0.4\textwidth]{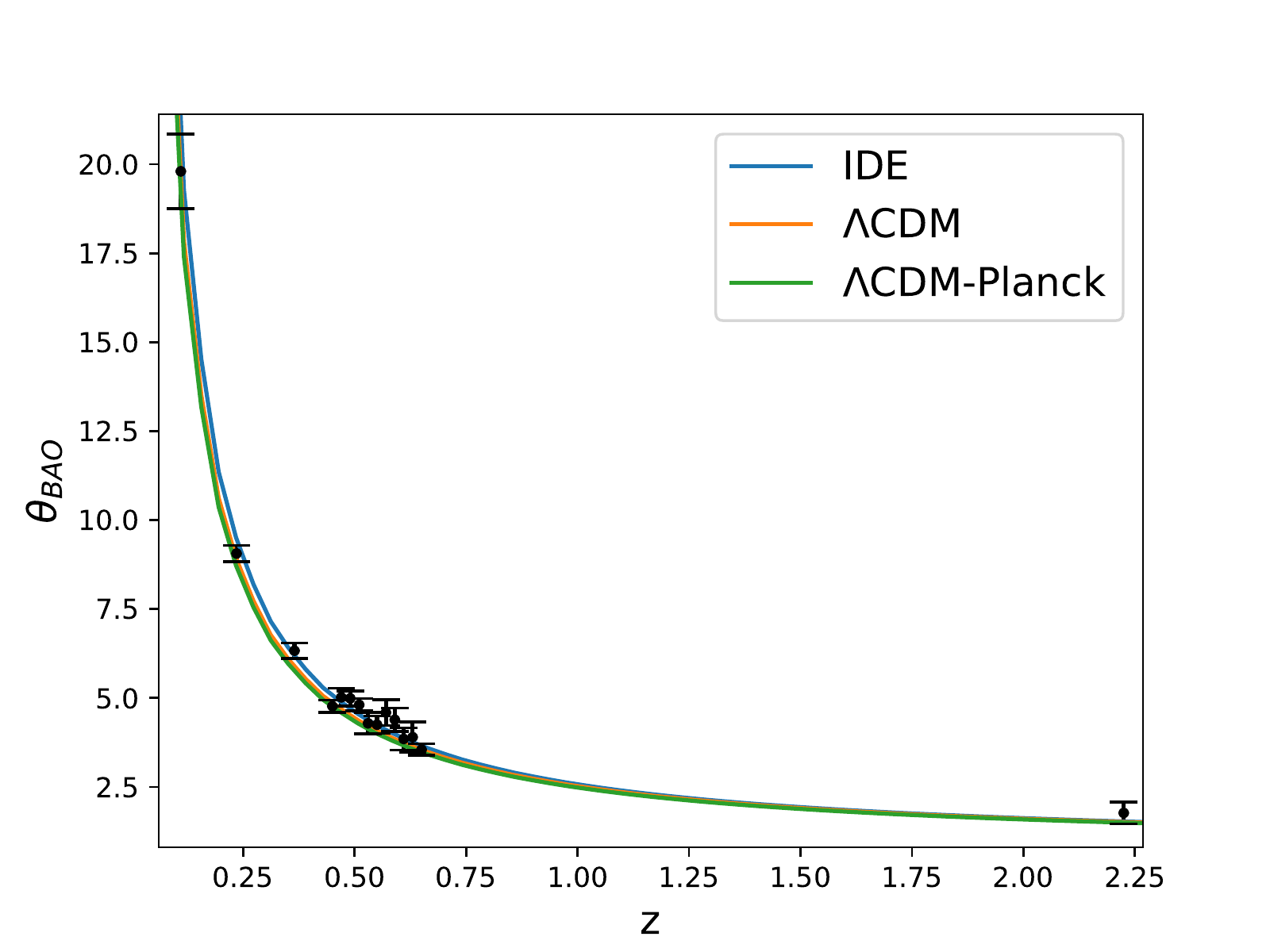} \,\,\,\,
\includegraphics[width=0.4\textwidth]{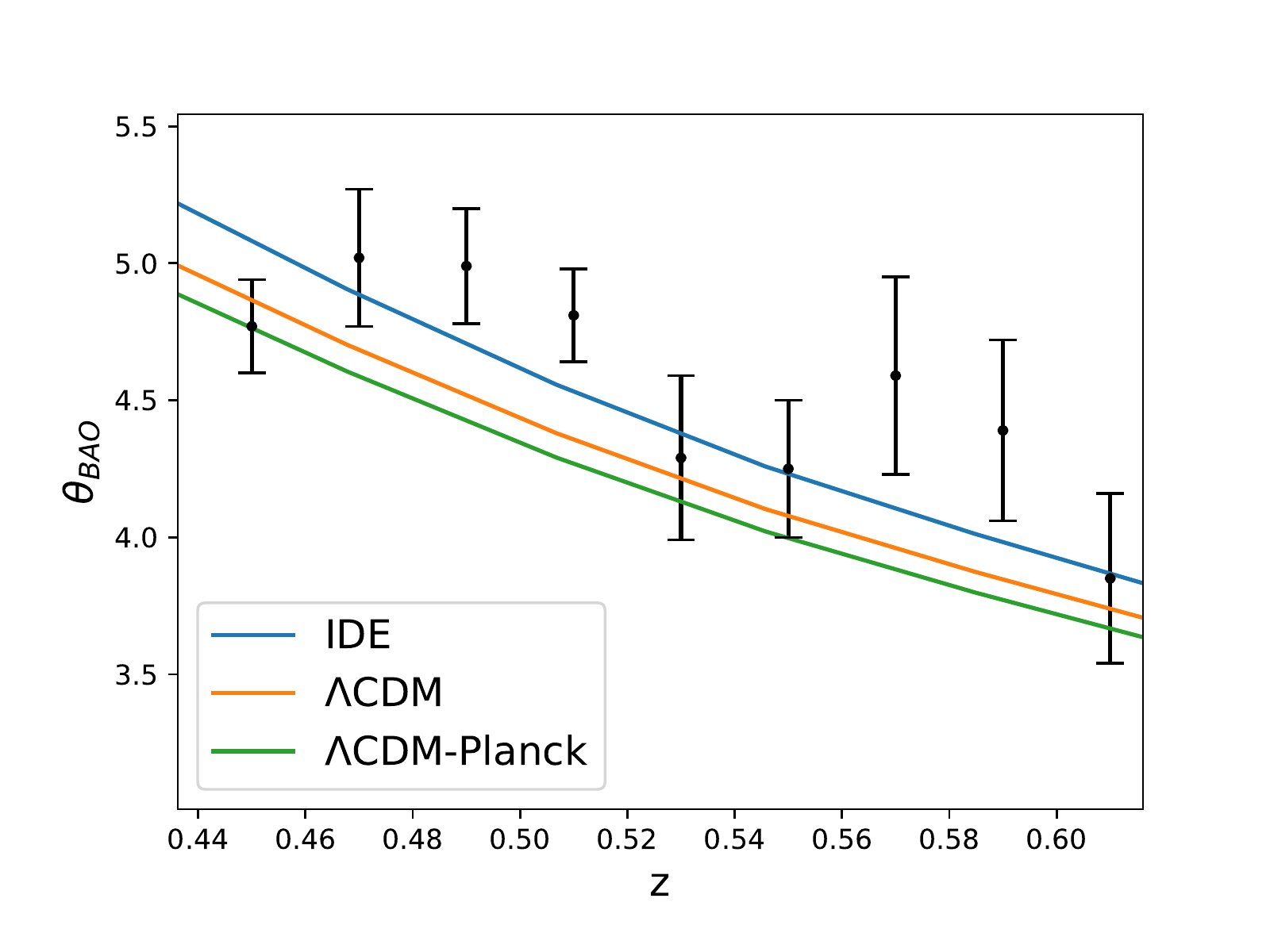}
\caption{Left panel: Best fit values for the $\Lambda$CDM and IDE models obtained from Planck + BAOtr joint analysis, and $\Lambda$CDM from best fit values from Planck only, compared against the BAOtr (i.e., $\theta_{\text{BAO}}$) sample. 
Right panel: Same as in the left panel, but for a specific $z$ range in the sample.}
\label{fig:th_best_fit}
\end{figure*}

In Table~\ref{tab.IDE}, we report the summary of the statistical analyses of the main parameters of interest considering the IDE model from several data combinations. In addition, we also show the constraints on the $\Lambda$CDM model using the same data sets in Table~\ref{tab.LCDM} in order to compare the results.

We first consider Planck data only and its combination with the lensing data. In these analyses, the dark coupling parameter $\xi$ is non-null at 1$\sigma$ confidence level (CL) from both the analyses. Also, it is well known that $\xi$ is strongly anti-correlated with $H_0$. Thus, we can notice a high value for $H_0$ when inferred from CMB analysis only, in such a way that by this perspective, this class of IDE models can alleviate the $H_0$ tension~\cite{DiValentino:2017iww,Kumar:2017dnp}. For these analyses, the Bayesian evidence is inconclusive, and the models, IDE and $\Lambda$CDM, cannot be statistically distinguished from each other.
When BAO data are added, we notice that the values of $H_0$ are pulled to low values totally compatible with Planck + BAO and Planck + BAO + lensing as predicted for the $\Lambda$CDM model. The parameter $\xi$ attains  lower bounds $\xi > -0.389, > -0.411$ at 95\% CL from Planck + BAO, Planck + BAO + lensing, respectively. Again, the statistical evidence is inconclusive and weak. The analysis with this BAO sample represents an update of the previous works.

Now, we consider the addition of transversal BAO sample. From both analyses, i.e., Planck + BAOtr and Planck + BAOtr + lensing, notably we find $\xi < 0$ at more than 3$\sigma$. The constraints on the Hubble constant read as 
$75.2^{+1.2}_{-0.75}$ km/s/Mpc at 68\% CL and $75.3^{+1.3}_{-0.75}$ km/s/Mpc at 68\% CL from Planck + BAOtr and Planck + BAOtr + lensing, respectively. These constraints are compatible with direct local measurements by the SH0ES team within 2$\sigma$. The Bayesian evidence for the IDE in this case is very strong. Here we can notice a clear difference when considering different BAO methodologies/samples to constrain the IDE framework. Since the $H_0$ values are now compatible with SH$0$ES measurements, ultimately, we perform a joint analysis considering the $H_0$ prior, that is equivalent to consider a $M_B$ prior for this model~\cite{Nunes:2021zzi}. In this final case, the constraint become slightly stronger than those reported without the $H_0$ prior, but with the same conclusion, i.e, a very strong evidence in favor of the IDE model. Figure~\ref{fig:PS_IDE} displays the parametric space on the plane $\xi$-$H_0$ at 68\%, 95\%, and 99\% CL for the analyses considered in this work.

To understand what happens in these constraints, in Figure~\ref{fig:th_best_fit} we show the theoretical prediction of $\theta_{\text{BAO}}(z)$ from the best-fit values of Planck + BAOtr for the IDE and $\Lambda$CDM models. The comoving sound horizon at the baryon drag epoch, $r_s$, is the main parameter which controls the amplitude of the angular scale $\theta_{\text{BAO}}(z)$, but as shown in Tables~\ref{tab.IDE} and~\ref{tab.LCDM}, for both models this scale is practically the same. This is well predicted and expected, as the IDE model does not modify the physics at early times. On the other hand, the possible theoretical presence for $\xi < 0$, will make the universe to expand faster at late times, and consequently, decrease the angular diameter distance in $\theta_{\text{BAO}}(z)$ prediction, making the amplitude of $\theta_{\text{BAO}}(z)$ to increase, without the need to change $r_s$ scale. Figure~\ref{fig:th_best_fit} on the right panel shows the measurements in the range $0.35 < z < 0.65$, where an excess in amplitude is noted in these measurements in relation to the Planck-$\Lambda$CDM best-fit.
Therefore, the IDE framework has the ability to fit better to the $\theta_{\text{BAO}}(z)$ estimates than $\Lambda$CDM model, as the $\theta_{\text{BAO}}(z)$ measurements have an excess in amplitude in relation to the Planck-$\Lambda$CDM best-fit prediction, where the $\Lambda$CDM model is not able to fully fit. On the other hand, even in $\Lambda$CDM framework, the Planck + BAOtr joint analysis constrains the Hubble constant to be $H_0 = 69.01 \pm 0.51$ km/s/Mpc at 1$\sigma$, which is higher than predicted by Planck and Planck + BAO analyses, and at the same time prefers a lower value for $S_8$, also helping with this tension. This difference occurs precisely to try to fit the excess in amplitude in the range $0.35 < z < 0.65$. However, this is not enough to fit all data points, as CMB data in combination with BAOtr prefer to keep low $H_0$ values in $\Lambda$CDM model, while for IDE the opposite is true. As already mentioned earlier, reducing the information from 3D to 2D increases the error bars on the measurements. These perspectives were investigated for $\Lambda$CDM and curvature in \cite{Nunes_Bernui_2020_bao} independently of CMB data and local distance ladder method. The 2D BAO sample allows higher values of $H_0$, despite that the 2D BAO data set alone does not have robustness to constrain the full parameter space of cosmological models. In addition to all details mentioned above, we know that in the IDE context, CMB data prefer higher $H_0$ values, and the 2D BAO data set also does the same. Thus, the combination Planck + BAOtr, also leads to higher $H_0$ values for this class of IDE, as a natural and expected behavior of this model. We reinforce that new 2D BAO measurements with more accuracy are necessary to investigate the $H_0$ tension, and to better constrain the cosmological models. Hence we can conclude that the difference between the addition of 3D or 2D BAO data is much less pronounced in $\Lambda$CDM than in extended models, questioning the reliability and robustness of BAO measurements in these cases, and the importance of their assumptions on the fiducial model in the data reconstruction process.

Further, we can notice that $S_8$ values are higher in the IDE model than in the $\Lambda$CDM (see Table \ref{tab.IDE} and Table \ref{tab.LCDM}). Also, the $S_8$ values in the IDE model are higher than the ones estimated by weak lensing and galaxy clustering surveys data \textit{assuming} the $\Lambda$CDM framework~\cite{KiDS:2020suj} (though they agree within the error bars~\cite{DiValentino:2019ffd}). However, this comparison is not relevant because $S_8$ value is model-dependent. In fact, the $S_8$ value estimated  by weak lensing and galaxy clustering surveys data should be compared with the one estimated by the other data (such as Planck) assuming the same underlying model. Furthermore, it is important to mention that the accurate modeling of IDE framework on weak lensing and galaxy clustering data, especially with regard to the dynamics on non-linear scales and its application on the related observables, has not yet been addressed in the literature. Therefore, more conclusive findings cannot be made regarding the $S_8$ tension in this class of IDE models. In conclusion, the IDE model considered in the present work, solves the $H_0$ tension, but leaves the $S_8$ tension open for further investigation.

\section{Final Remarks}
\label{sec:conclusions}

The main aim of this work is to answer the question: What is the impact of the BAO measurements besides the usual traditional approach of the 3D BAO sample on models beyond the $\Lambda$CDM model, when applied on models that can have significantly different dynamical behavior with respect to the $\Lambda$CDM prediction? In this regard, we have used transversal BAO measurements, which are weakly dependent on a cosmological model  to obtain new observational constraints on IDE cosmologies. For a specific class of models, we find that 3D BAO and 2D BAO measurements can generate very different observational constraints on the coupling parameter quantifying an interaction between DE and DM. This difference is much more significant that the one obtained by comparing the constraints of the $\Lambda$CDM model using 3D BAO and 2D BAO data. We actually find a very strong evidence for IDE model when 2D BAO are taken into account. Moreover, unlike Planck + BAO joint analysis, where it is not possible to solve the $H_0$ tension, we note that the joint analysis Planck + BAOtr is in perfect agreement with the direct local distance ladder measurements by the SH0ES team. A different transversal BAO sample of the adopted here is presented in \cite{Menote_2022}. We check that, in principle, this sample does not change our main conclusions. Therefore, we conclude that the minimal baseline Planck + BAOtr in the context of IDE framework solves the Hubble constant tension. 


\begin{acknowledgments}
\noindent  

The authors thank the referees for useful comments. AB thanks a CNPq fellowship. EDV is supported by a Royal Society Dorothy Hodgkin Research Fellowship. SK gratefully acknowledges support from the Science and Engineering Research Board (SERB), Govt. of India (File No.~CRG/2021/004658). 
This article is based upon work from COST Action CA21136 Addressing observational tensions in cosmology with systematics and fundamental physics (CosmoVerse) supported by COST (European Cooperation in Science and Technology). RCN thanks the
CNPq for partial financial support under the project No.
304306/2022-3.
\end{acknowledgments}

\bibliography{PRD}

\end{document}